\documentclass{iau} 
\usepackage{graphicx}
\usepackage{natbib}

\title[IAUS 356~~AGN fueling and feedback, from pc to kpc scale] 
{AGN fueling and feedback}

\author[Francoise Combes]   
{Francoise Combes}

\affiliation{Observatoire de Paris, LERMA, Coll\`ege de France, CNRS, \\PSL University, 
Sorbonne University, UPMC, Paris \\ email: {\tt francoise.combes@obspm.fr} }

\pubyear{2019}
\volume{356}  
\jname{Nuclear Activity in Galaxies Across Cosmic Time}
\editors{M. Povic et al.}
\begin{document}

\maketitle

\begin{abstract}
Dynamical mechanisms are essential to exchange angular
momentum in galaxies, drive the gas to the center, and fuel the central
super-massive black holes. While at 100pc scale, the gas is sometimes
stalled in nuclear rings, recent observations reaching $\sim$10pc scale
have revealed, within the sphere of influence of the black hole, smoking
gun evidence of fueling.
Observations of AGN feedback are described, together
with the suspected responsible mechanisms. Molecular outflows are frequently
detected in active galaxies with ALMA and NOEMA, with loading factors
between 1 and 5. When driven by AGN with escape velocity,
these outflows are therefore a clear way to moderate or suppress star formation.
Molecular disks, or tori, are detected at 10pc-scale, kinematically decoupled
from their host disk, with random orientation. They can be used to measure
the black hole mass.
\end{abstract}

\firstsection 
\section{Introduction: the main paradigm}

For a long time, the main unification paradigm to explain the
large variety of AGN, and in particular the type 1 with broad lines (BLR),
and the type 2, with only narrow lines (NLR), has been the obscuration 
of the accretion disk by a dusty torus, hidding the BLR and showing only
the NLR to the observer \citep{Urry1995}.
This idea is still valid for a certain number
of AGN (for instance type 2 Seyfert which reveal their BLR in polarized light),
but is known not to be the only parameter distinguishing the various AGN,
since there exist intrinsic differences between Sy1 and Sy2, and also
a number of changing look AGNs have been discovered, while their
transformation from Sy1 to Sy2 an vice-versa in time-scales of dozens of years
has nothing to do with obscuration \citep{Denney2014, McElroy2016}.

In the last decade, high spatial resolution observations
 in the mid-infrared  with the VLT Interferometer (VLTI) showed
that the dust on parsec scales is not mainly in a thick torus,
but instead in a polar structure,  forming like a hollow cone, perpendicular
to a thin disk \citep[e.g.,][]{Asmus2016, Hoenig2019} and references therein).
 A large majority of objects ($\sim$ 90\%) reveal that most ($>$ 50\%)
 of the mid-infrared emission of the dust comes from a polar structure, 
 leaving little room for a torus contribution \citep{Asmus2019}.
The new view which is sketched now for the cold gas is a two component medium,
one inflowing in a thin disk, where millimeter lines have been found, 
with also H$_2$O masers, or the ro-vibrational lines of warm H$_2$, and an 
outflowing component, in the perpendicular direction,
responsible for the polar dust emission. Molecular outflows will be driven along 
the borders of this hollow cone. The inner boundary of the dusty thin disk 
would correspond to the sublimation of
the dust by the AGN radiation. The circum-nuclear molecular disk 
could be observed  to extend
down to smaller distances from the center.

In the following, I review recent observations at high-angular
resolution of the molecular gas, showing how gravity torques
can help feeding the central black hole, through exchange of
angular momentum.
These are the consequences of nuclear 
dynamical features, such as nuclear bars \& spirals.
 The same observations might reveal outflows, occuring
 simultaneously with inflow, albeit in different directions.
 These outflows are due to the feedback effect of the AGN,
 added to the star formation feedback.
 Frequently in nearby active galaxies, molecular circum-nuclear disks
 are observed; with decoupled kinematics from the larger-scale
 galactic disks. These parsec-scale structures may be identified to
 molecular tori, able to obscure the central accretion disks.
 Being within the sphere of influence of the black holes,
 they help to determine their mass.

\begin{figure}[t]
\begin{center}
 \includegraphics[width=0.99\textwidth]{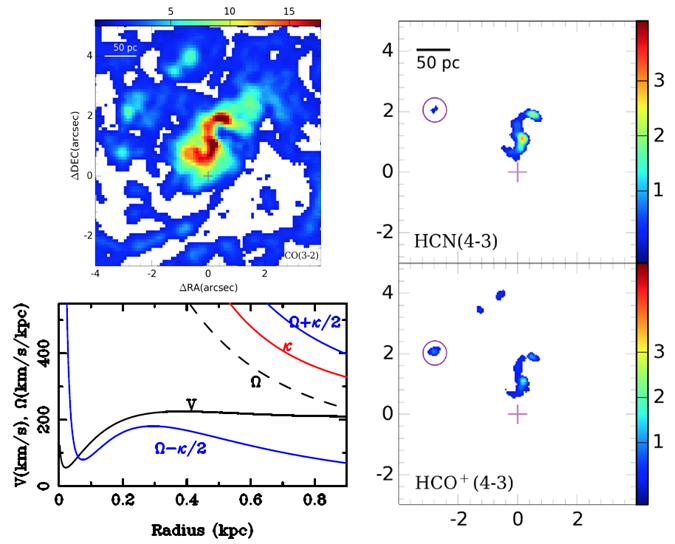} 
 \caption{ALMA observations of the Seyfert-2 galaxy NGC~1808. Left is a zoomed
   4''x4'' region of the CO(3-2) intensity map, showing the nuclear trailing spiral.
   Right are the intensity maps of the dense gas tracers HCN(4-3) and HCO$^+$(4-3). The size
   is given in parsec  by the 50pc bar. At the bottom left, is the model rotation curve,
   taken from both the near-infrared (potential from old stars) and the gas observed velocity
   \citep{Salak2016}. From Audibert et al. (2020) in prep., and
\cite{Combes2019}.}
   \label{fig1}
\end{center}
\end{figure}

\section{Nuclear trailing spirals}

Observed nuclear spirals bring clear evidence of fueling. They 
were detected first in NGC~1566 \citep{Combes2014}, during the first 
ALMA observations, with 0.5 arcsecond  $\sim$25pc resolution.
In this barred spiral, there is an r=400pc ring, corresponding to the inner
Lindblad resonance (ILR) of the bar. The stellar periodic orbits, which
attract all regular orbits, change by 90$^\circ$ at each resonance
\citep{Contopoulos1980}. They are parallel to the bar inside corotation,
and become perpendicular in-between the two ILRs.

The gas behaviour may be derived from these orbits; the gas elliptical streamlines
tend to follow them, but gas clouds are subject to collisions. The ellipses
are gradually tilted by 90$^\circ$ and wind up in spiral structures. The precession
rate of these elliptical orbits in the epicyclic approximation is 
equal to $\Omega-\kappa$/2, with $\Omega$ the rotation frequency = V/r, and
$\kappa$ being the epicyclic frequency. When the dissipative gas is driven
to the center, the precession rate first increases, since $\Omega-\kappa$/2
increases, and the spiral is trailing. But inside the ILR, the precession
rate reaches a maximum and declines, and this changes the winding
sign of the gas, which is expected to be in a leading spiral.
 The reversal of the winding sense has also the consequence of reversing
 the sign of the gravity torques exerted by the stellar bar on the gas
 \citep{Buta1996}. The torque is negative from corotation to ILR,
 but then is positive inside the ILR, this confines the gas
 in the ILR ring, where it forms stars actively.

 However, in the presence of the central black hole, all frequencies
 $\Omega$ and $\Omega-\kappa$/2 increase again towards the center,
 as shown in Fig. \ref{fig1}. This is able to reverse the winding sense
 of the gas spiral, if a sufficient amount of gas is within the influence 
 of the black hole. In that case, the gravity torques become negative again,
 and the gas is driven to the center, to fuel the AGN. The very
 presence of a nuclear trailing spiral is smoking gun evidence of 
 the AGN fueling.

 Such trailing nuclear spirals have been found also in NGC~613
 \citep{Audibert2019} and in NGC~1808 (Audibert et al. 2020, in prep.),
 as can be seen in Fig. \ref{fig1}. The nuclear spiral is conspicuous
 in the CO(3-2) emission line, but also in the dense gas tracers like 
 HCN, HCO$^+$ or CS. The rotation curves of these galaxies, derived 
 both from the stellar potential (traced by near-infrared images)
 and the observed kinematics of the gas, confirm that the nuclear
 gas is falling within the sphere of influence of the black holes.

 The gravity torques have been measured on the deprojected images
 of the molecular gas, with the method described in \cite{Santi2012},
 and the torques are indeed negative \citep{Audibert2019}.
 The computation allows to quantify the strength of the torques,
 and shows that the gas loses most of its angular momentum in one rotation,
 i.e. 10 Myr.

\section{Molecular outflows}

In some cases, molecular outflows are observed simultaneously
with the evidence of fueling. This is the case of NGC~613, 
where a very short (23pc) and small velocity (300km/s)
outflow is detected on the
minor axis, in the same direction of the radio jet, mapped at cm
wavelength with the VLA \citep{Audibert2019}. 
This was possible thanks to the ALMA resolution of 60 mas (5pc).
The coincidence 
with the radio jet strongly suggests that the outflow is AGN
feedback in the radio mode. 

In other cases (e.g. NGC~1566) no 
molecular outflow is detected. In NGC~1808, an outflow has been
known for a long time, at large scale, from ionized gas and dust
ejected perpendicular to the plane, creating an extra-planar medium
\citep{Busch2017}.
 However, at parsec scale, the CO emission from ALMA does not reveal
 any outflow. In that case, it can be concluded that the outflow
 is driven by supernovae feedback, but not from the AGN.

In that NGC~1808 galaxy, the starburst and AGN contributions can be
distinguished by the 
diagnostics of line ratios between HCN and HCO$^+$ or CS.
Close to the AGN, the HCN line is considerably enhanced 
\citep{Usero2004, Krips2008}, due to the X-rays from the AGN.

\bigskip

In these nearby low-luminosity AGN, the main mechanism
to drive molecular outflows is entrainement by the radio jets.
 There are two main modes identified for AGN feedback:
\begin{itemize}
\item the quasar mode, or radiative mode with winds. This occurs
when the AGN luminosity is close to the Eddington luminosity, 
for young quasars at high redshift essentially. Then the
radiation pressure exerted on the ionized gas (with the 
Thomson cross-section) can drive an ionized wind.
A similar effect can occur with the radiation pressure on dust,
with a higher cross-section.

\item the radio mode, or kinetic mode, due to radio jets. This
occurs when the AGN luminosity is lower than 1/100~th of the
Eddington luminosity, typically at low redshift.
Massive galaxies, and early-type galaxies are frequently the host
of radio-loud AGN. These low-luminosity AGN are radiatively 
inefficient (ADAF).
 In the low redshift universe, strong AGN feedback
in the radio mode is observed frequently in cooling flows of
galaxy clusters,
\end{itemize}

\bigskip

There is a radio-mode molecular outflow in the proto-typical
Seyfert 2 galaxy NGC~1068. The nucleus is
off-centered, and the radio jet is not perpendicular to the plane;
therefore it is sweeping out some gas in the disk \citep{Santi2014}. The
molecular outflow is estimated at 63M$_\odot$/yr, or
 10 times the star formation rate in the central region.

In NGC~1068, the high resolution of ALMA has 
allowed the detection of a molecular torus, with both
the CO(6-5) emission line, and with the dust emission
at 432$\mu$m. The radius of the torus depends on the tracer,
it is with CO(6-5) 5-6 pc, but larger in low-J lines, like CO(2-1).
 The various tracers yield different aspects of the cold medium
 in the center. The CO disk is warped, and appears more 
inclined than the H$_2$O maser disk \citep{Santi2016}.
It is possible that the circum-nuclear
disk is unstable, through the Papaloizou-Pringle instability
\citep{Papaloizou1984}.

The molecular torus is located just 
at the base of a polar dusty cone. The cone
has been mapped in the near-infrared, with SPHERE on
the VLT, and in particular the polarisation is revealing
beautifully the conical struture \citep{Gratadour2015}.

\begin{figure}[t]
\begin{center}
 \includegraphics[width=0.99\textwidth]{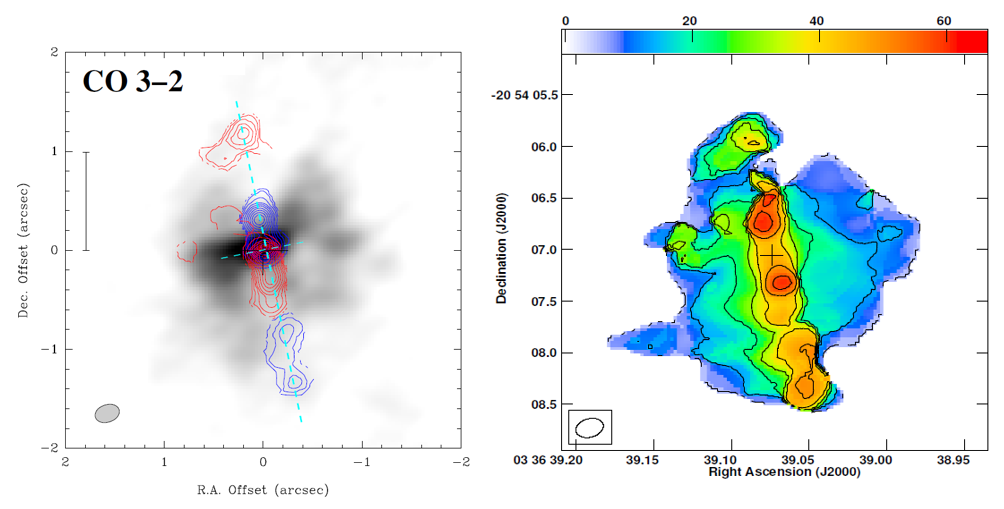} 
 \caption{ALMA CO(3-2) map of the galaxy NGC~1377: Left, the grey-scale corresponds
   to the emission close to systemic velocity, while he high velocity (80 to 150km/s from
   V$_{sys}$) is indicated as red and blue for red-shifted and blue-shifted
   emission respectively. The vertical bar indicates a scale of 100 pc.  The dashed line
   indicates the jet axis (and nuclear disk) orientations.
   Right: the velocity dispersion (moment 2) map, from 0 to 66 km/s. The cross indicates
   the position of the continuum peak at 345 GHz.
   From \cite{Aalto2016}.}
   \label{fig2}
\end{center}
\end{figure}

\bigskip

The radio mode might also be at play in the lenticular galaxy NGC~1377,
although no radio jet has been detected, and the galaxy is the most
radio-quiet found, in terms of the radio-far-infrared correlation.
A weak radio emission is detected at the center, but much weaker than 
expected from the well known correlation \citep{Helou1985}.

As shown in Fig. \ref{fig2}, ALMA has detected in this galaxy a very narrow
molecular outflow \citep{Aalto2016}. The molecular outflow changes sign
along the flow, on each side of the galaxy. This very surprising
behaviour is very rare, and is interpreted as the precession of the jet, 
which happens to be very little inclined on the sky plane.
Therefore a precession of 10$^\circ$ only is sufficient to reverse
the sign of the flow velocity towards the observer. Such precession is
observed in micro-quasars jets in the Milky Way, for instance SS433
\citep{Mioduszewski2005}.

The molecular gas in the cone is derived to be  
10$^8$ M$_\odot$, and there is 10$^7$ M$_\odot$
in the outflow \citep{Aalto2016}. A model of a precessing 
molecular outflow has been found compatible with the data.
The velocity of the flow, according to the model, 
ranges between 250 and 600km/s. 
The flow is launched at a distance from the center lower than 10pc.
It cannot be driven by supernovae feedback, since there is no starburst in
the galaxy, and the flow would not be so collimated. A radio jet
must exist at a low level, or has existed in a recent past.

\begin{figure}[t]
\begin{center}
 \includegraphics[width=0.99\textwidth]{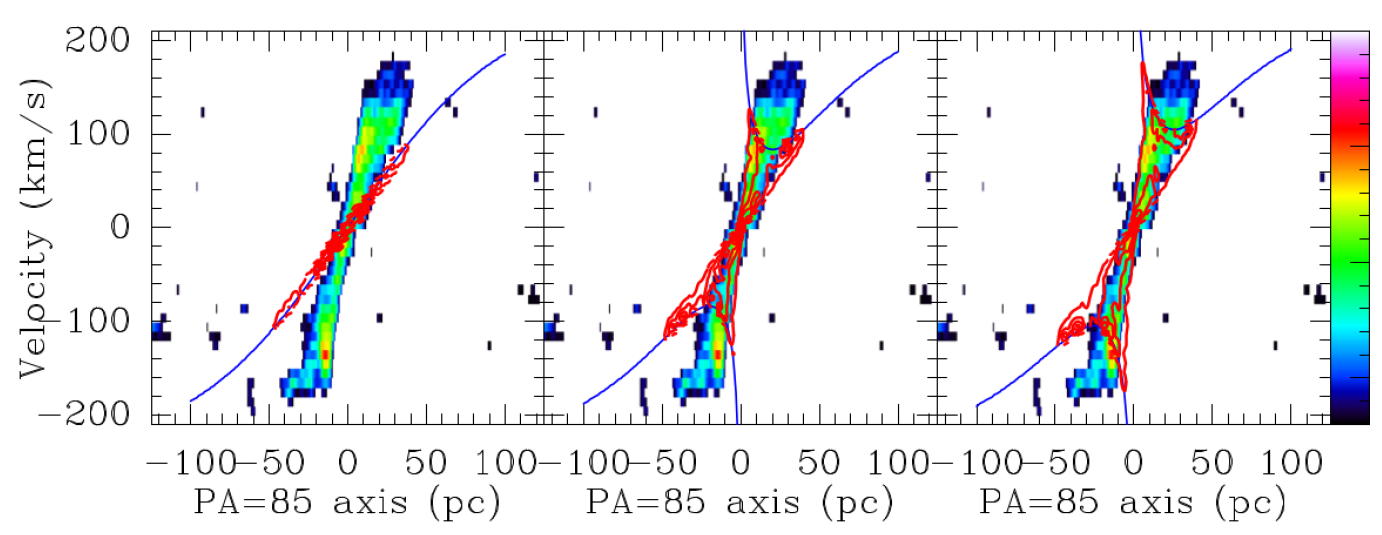} 
 \caption{ALMA CO(3-2) observations of the galaxy NGC~1672. The 3 panels
   are the position -velocity diagrams in color, along the major axis of the molecular disk
   (or torus), with superposed in red, three models with different values of the central
   black hole mass (0, 2.5 and 5.0 10$^7$ M$_\odot$. The predicted circular velocity is
   reproduced in blue lines. From \cite{Combes2019}.}
   \label{fig3}
\end{center}
\end{figure}

\section{Molecular tori}

When the high-angular resolution is available with ALMA,
the frequency of detection of molecular tori in nearby
Seyfert galaxies is quite high, 7 out of 8, as shown in
the Table \ref{tab1}. 

\begin{table}
{\centering
	\caption[]{Radii, masses, and inclinations of the molecular tori}
\begin{tabular}{lccccc}
\hline
        Galaxy&Radius&M(H$_2$)  &inc($^\circ$)&PA($^\circ$)&inc($^\circ$)\\
& (pc)   & 10$^7$ M$_\odot$  & torus&torus& gal \\
\hline
NGC~613   &  14$\pm$3   & 3.9$\pm$1.4  & 46$\pm$7 &0$\pm$8   &36\\
NGC~1326  &  21$\pm$5   & 0.95$\pm$0.1 & 60$\pm$5 &90$\pm$10 &53\\
NGC~1365  &  26$\pm$3   & 0.74$\pm$0.2 & 27$\pm$10&70$\pm$10 &63\\
NGC~1433  &  --         & --           & --       & --       &67\\
NGC~1566  &  24$\pm$5   & 0.88$\pm$0.1 & 12$\pm$12&30$\pm$10 &48\\
NGC~1672  &  27$\pm$7   & 2.5$\pm$0.3  & 66$\pm$5 &0$\pm$10  &28\\
NGC~1808  &  6$\pm$2    & 0.94$\pm$0.1 & 64$\pm$7 &65$\pm$8  &84\\
NGC~1068  &  7$\pm$3    & 0.04$\pm$0.01& 80$\pm$7 &120$\pm$8  &24\\
\hline
\end{tabular}
 \label{tab1}\par
	}
\end{table} 

We call ''molecular torus" the circum-nuclear molecular disk, of parsec 
scale, which is kinematically and spatially decoupled from the rest of the 
disk. It might not appear as a torus, except in favorable cases, like in 
NGC~1365. The molecular tori are located within the sphere
of influence of the black holes, and can serve to measure their mass.
 The example of the Sy2 NGC~1672, where the torus is seen almost edge-on,
 is displayed in Fig. \ref{fig3}. The ALMA resolution is 3pc.
 The position-velocity diagrm reveals a strong velocity gradient
 of about 180 km/s in 30pc.

\bigskip

A 3D modelisation of the dynamics of the nuclear disk has led to
 a black hole mass of 5 10$^7$ M$_\odot$ \citep{Combes2019}.
The black hole mass determination from the gas kinematics is a
precious method, for these low-luminosity AGN, which are late-type 
spiral galaxies frequently with a pseudo-bulge.
For them the scaling relation between the central velocity
dispersion and the mass of their black hole is quite scattered
\citep[e.g.,][]{Graham2011}. This is not the case of more massive
early-type galaxies, where the gas method has also been used
with success \citep{Davis2018}.

\begin{figure}[t]
\begin{center}
 \includegraphics[width=0.99\textwidth]{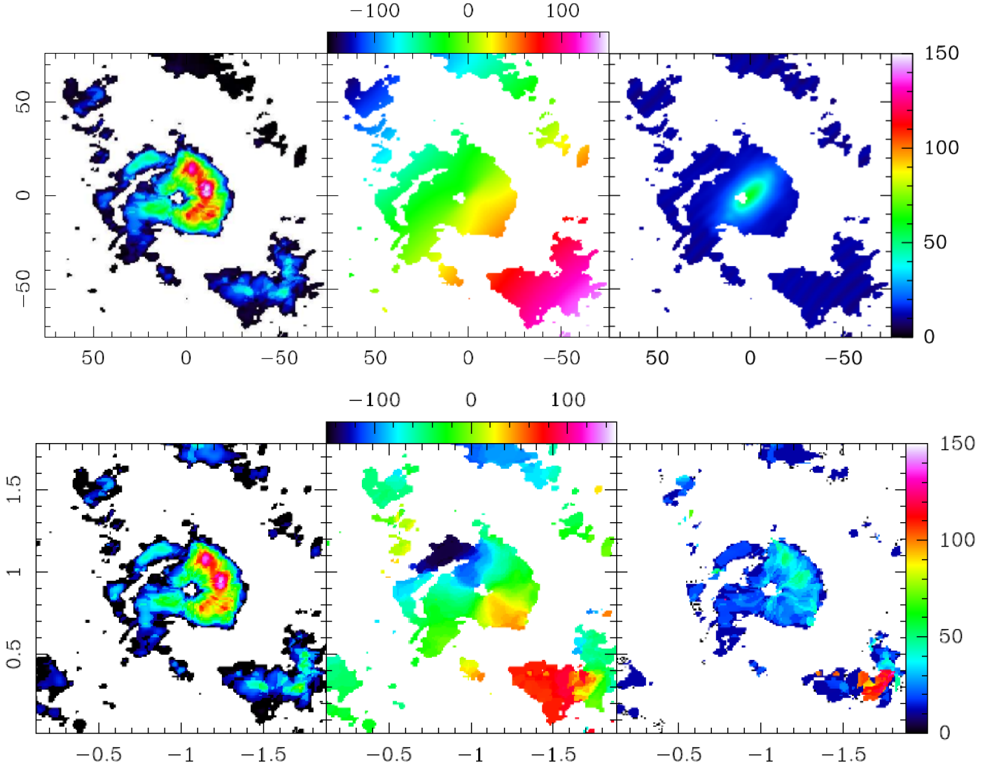} 
 \caption{ALMA CO(3-2) observations of the galaxy NGC~1365.
   In each row, the three panels display the three first moments
   (intensity, velocity and dispersion). 
   The velocity color scale is in km/s.  
The model correspnds to a black hole mass of 4 10$^6$ M$_\odot$.    
Top is the model, and bottom the observations.  
	The scale is spatial offset in parsec
	for the model, and in arcsec for the observations..
	From \cite{Combes2019}.}
   \label{fig4}
\end{center}
\end{figure}

The 3D modelisation of the molecular torus dynamics for the
Sy 1.8 galaxy NGC~1365 is displayed in Fig. \ref{fig4}.
The black hole mass is 4 10$^6$ M$_\odot$ \citep{Combes2019}.

One common feature to the detected molecular tori, is their random
orientation and decoupling with the large-scale disk of their host,
cf Table \ref{tab1}.
This feature is not unexpected, given the very different time-scales
at parsec scale and kpc-scales, and the almost spherical potential
of the central galaxy at small scale.  
The material near the black hole will eventually inflow and fuel the AGN,
and will be replaced by accreted gas, coming with different orientations
and angular momenta. Numerical simulations show examples of
central starbursts, where gas is ejected in a fountain through
supernovae feedback, and may cool and fall back with random orientation,
sometimes in a polar ring \citep{Renaud2015, Emsellem2015}.

 Examples of nuclear disks decoupled from their host disks
 are frequently seen, as in the Milky-Way, where a
 circum-nuclear ring of 2-3pc in radius surrounds an
 almost face-on mini-spiral, or in NGC~4258, where
 anomalous arms (in fact the radio jet) are winding
 in the plane of the large-scale disks, with the normal
 spiral arms.

 In the edge-on disc galaxy HE1353-1917,
 \cite{Husemann2019} have found a radio jet 
impinging the molecular gas of the host, and producing an outflow;
in addition, [OIII] emission in a cone oriented at
only $\sim$ 10$^\circ$ from the edge-on plane, reveals
gas illuminated by the AGN, collimated by a highly inclined torus.

\section{Summary}

Recent high spatial resolution observations with ALMA have revealed
the role played by dynamical features like bars to drive
the gas at parsec scales, and fuel the AGN. The process is
occurring in several steps, first from corotation, the gas is
driven inwards, and piles up in a ring at the inner Lindblad 
resonance, at 100~pc-scale. Then, either through a nuclear bar,
or through dissipation, the gas may be driven further in; 
there, within the sphere of influence of the black hole,
it rapidly loses its angular momentum, settles in
a 10pc-scale disk or torus and fuels the AGN.

Simultaneously, the AGN feedback through radiative or radio mode,
according to the Eddington ratio, may drive molecular outflows,
in a perpendicular direction. In some cases, the feedback is
from the supernovae of a central starburst, in association (or not)
to the AGN feedback. The AGN feedback can have a strong coupling
with the gas in the host disk, due to the mis-alignment
of the nuclear and large-scale disks.

Circum-nuclear disk, or molecular tori are frequently
detected in nearby Seyferts, with a random orientation,
kinematically decoupled from their host disk. This  
mis-alignment between small scales and large scales is
due to random gas accretion, and different dynamical time-scales,

\end{document}